\documentclass[prl,twocolumn,showpacs,amsmath,amssymb,superscriptaddress]{revtex4-1}
\usepackage{graphicx}
\usepackage{bm}
\usepackage[normalem]{ulem}
\usepackage{color}
\usepackage{amsfonts}
\usepackage{amssymb}
\usepackage{amsmath}
\usepackage{mathtools}
\usepackage[mathscr]{euscript}
\usepackage{gensymb}
\usepackage{epsfig}

\graphicspath{{./}}

\newcommand{\T}{{\mathcal{T}}}
\newcommand{\TE}{{\textbf{\textsf{T}}}}
\newcommand{\G}{{\textbf{\textsf{G}}}}

\newcommand{\kp}{{\textbf{k}_{||}}}
\newcommand{\1}{{\mid}}
\newcommand{\OS}{{CoPt$\1$MgO$\1$CoPt }}
\newcommand{\OSNS}{{CoPt$\1$MgO$\1$CoPt}}
\newcommand{\AS}{{CoPt$\1$MgO$\1$Pt }}
\newcommand{\ASNS}{{CoPt$\1$MgO$\1$Pt}}

\begin{document}

\preprint{APS/123-QED}

\title{Large Tunneling Anisotropic Magneto-Seebeck Effect in a \AS Tunnel Junction}

\author{V. P. Amin}
\thanks{aminvp@physics.tamu.edu}
\affiliation{Department of Physics, Texas A\&M University, College Station, Texas 77843-4242, USA}
\author{J. Zemen}
\affiliation{Department of Physics, Blackett Laboratory, Imperial College London, London SW7 2AZ, United Kingdom}
\author{J. \v{Z}elezn\'y}
\affiliation{Institute of Physics ASCR, v.v.i., Cukrovarnick\'a 10, 162 53 Praha 6, Czech Republic}
\affiliation{Faculty of Mathematics and Physics, Charles University in Prague, Ke Karlovu 3, 121 16 Prague 2, Czech Republic}
\author{T. Jungwirth}
\affiliation{Institute of Physics ASCR, v.v.i., Cukrovarnick\'a 10, 162 53 Praha 6, Czech Republic}
\affiliation{School of Physics and Astronomy, University of Nottingham, Nottingham NG7 2RD, UK}
\author{Jairo Sinova}
\affiliation{Institut f\"ur Physik, Johannes Gutenberg-Universit\"at Mainz, 55128 Mainz, Germany}
\affiliation{Department of Physics, Texas A\&M University, College Station, Texas 77843-4242, USA}
\affiliation{Institute of Physics ASCR, v.v.i., Cukrovarnick\'a 10, 162 53 Praha 6, Czech Republic}

\date{\today}

%%%%%%%%%%%%%%%%%%%%%%%%%%%%%%%%%%%%%%%%%%%%%%%%%%%%%%
% Abstract
%%%%%%%%%%%%%%%%%%%%%%%%%%%%%%%%%%%%%%%%%%%%%%%%%%%%%%

\begin{abstract}

We theoretically investigate the \emph{Tunneling Anisotropic Magneto-Seebeck effect} in a realistically-modeled \AS tunnel junction using coherent transport calculations.  For comparison we study the tunneling magneto-Seebeck effect in \OS as well.  We find that the magneto-Seebeck ratio of \AS exceeds that of \OS for small barrier thicknesses, reaching $175\%$ at room temperature.  This result provides a sharp contrast to the magnetoresistance, which behaves oppositely for all barrier thicknesses and differs by one order of magnitude between devices.  Here the magnetoresistance results from differences in transmission brought upon by changing the tunnel junction's magnetization configuration.  The magneto-Seebeck effect results from variations in asymmetry of the energy-dependent transmission instead.  We report that this difference in origin allows for \AS to possess strong \emph{thermal} magnetic-transport anisotropy.

\end{abstract}

\maketitle

%%%%%%%%%%%%%%%%%%%%%%%%%%%%%%%%%%%%%%%%%%%%%%%%%%%%%%
% Begin Main Body
%%%%%%%%%%%%%%%%%%%%%%%%%%%%%%%%%%%%%%%%%%%%%%%%%%%%%%

\begin{figure}[h!]
\center{\includegraphics[width=1\linewidth, trim = 5pt 0pt 5pt 0pt, clip]{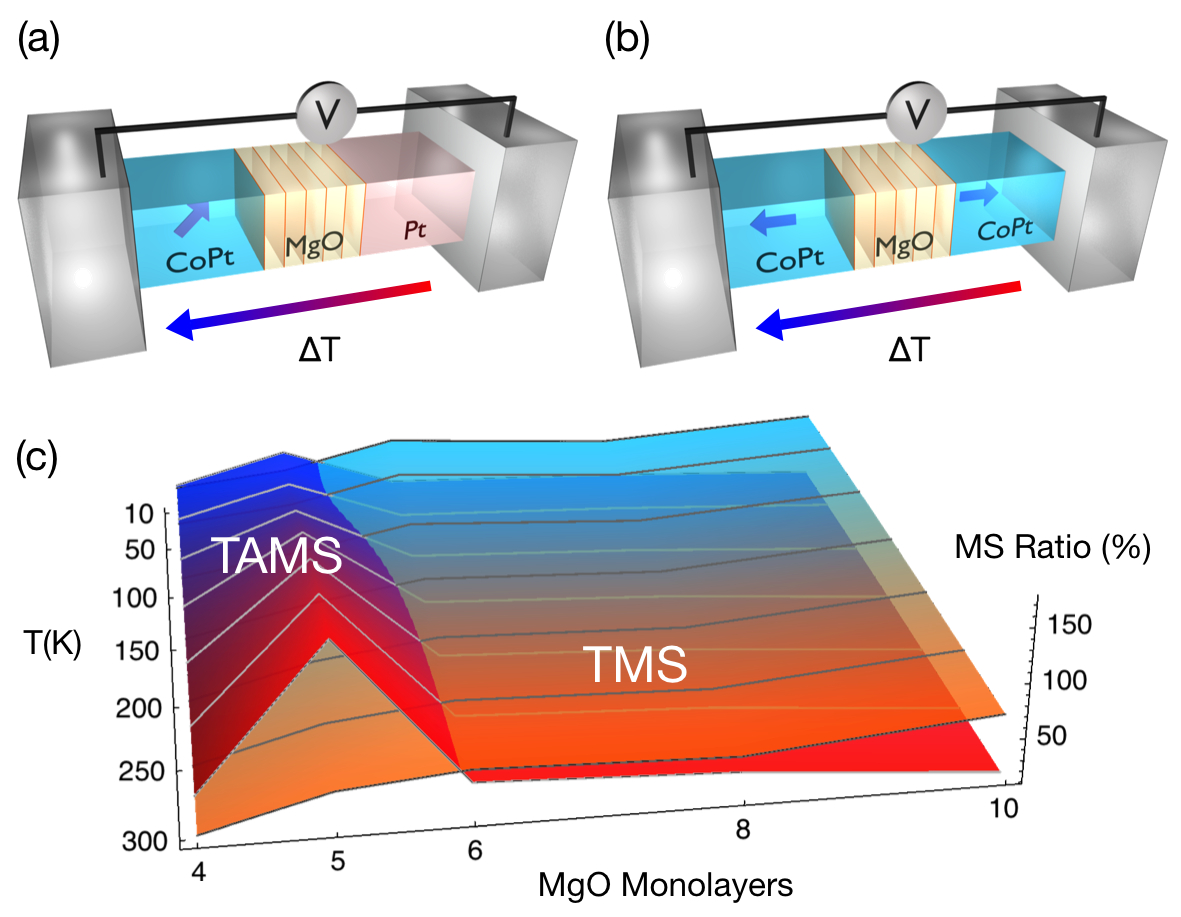}}
\caption{
\textbf{(a)}
The anisotropic MTJ (\ASNS).  A temperature gradient induces an open-circuit voltage across the contacts, known as the Seebeck effect.  Rotating the magnetization of CoPt produces varying Seebeck coefficients, also called the \emph{Tunneling Anisotropic Magneto-Seebeck} (TAMS) effect.
\textbf{(b)}
The normal MTJ (\OSNS), which exhibits the \emph{Tunneling Magneto-Seebeck} (TMS) effect.
\textbf{(c)}
The TAMS and TMS ratios plotted versus temperature and barrier thickness. The TAMS ratio (blue/red) surpasses the TMS ratio (cyan/orange) at small barrier thicknesses at all temperatures.  This provides a contrast to the behavior of the TMR and TAMR ratios (discussed in Fig. \ref{fig:MRResults}).
}
\label{fig:DeviceModel}
\end{figure}

Due to their presence in hard-disk drives and growing potential as commercially viable memory bits, magnetic tunnel junctions (MTJs) continue to provide impetus for scientific study.  The demand for smaller devices and efficient energy consumption mandates further investigation of their thermal properties.  Such considerations recently prompted a renewed interest in the long-known Seebeck effect, in which a temperature gradient spanning a material induces a voltage.  The discovery of corresponding thermal effects in spin-polarized systems heralded a new field of research known as Spin Caloritronics \cite{BauerSpinCaloritronics}.  In one such effect the (charge) Seebeck Coefficient changes as a function of a device's magnetization configuration - known as the magneto-Seebeck effect or magnetothermopower - in analogy with the magnetoresistance.  Recently observed experimentally \cite{WalterMagnetoSeebeck, LiebingMagnetothermopower, NaydenovaThermopower, TMSExperiment1, TMSExperiment2, TAMSExperiment1, TAMSExperiment2} and studied theoretically \cite{MagnetoSeebeckTheoryAbInitio1, MagnetoSeebeckTheoryAbInitio2, MagnetoSeebeckTheoryBoltzmann2}, the magneto-Seebeck effect enables one to tune the thermal properties of an MTJ via magnetic field, potentially enabling thermal spin-logic or assisting in the recycling of wasted heat.  We numerically study two devices: \AS (which we call an \emph{anisotropic} MTJ) and \OS (a \emph{normal} MTJ).  Although the magnetoresistance ratios of both devices differ by one order of magnitude, we find that their magneto-Seebeck ratios are comparable.  Furthermore, the anisotropic MTJ (or AMTJ for short) produces magneto-Seebeck ratios exceeding those of the normal MTJ at small barrier widths (shown in Figs.\ \ref{fig:DeviceModel} and \ref{fig:SeebeckandMSResults}), peaking at values of $68\%$ at $0$K and $175\%$ at $300$K. 

%%%%%%%%%%%%%%%%%%%%%%%%%%%%%%%%%%%%%%%%%%%%%%%%%%%%%%
% Figure Device Model and Main Result
%%%%%%%%%%%%%%%%%%%%%%%%%%%%%%%%%%%%%%%%%%%%%%%%%%%%%%
%
%
%%%%%%%%%%%%%%%%%%%%%%%%%%%%%%%%%%%%%%%%%%%%%%%%%%%%%%
% End Figure Device Model and Main Result
%%%%%%%%%%%%%%%%%%%%%%%%%%%%%%%%%%%%%%%%%%%%%%%%%%%%%%

%%%%%%%%%%%%%%%%%%%%%%%%%%%%%%%%%%%%%%%%%%%%%%%%%%%%%%
% Figure Theoretical Model
%%%%%%%%%%%%%%%%%%%%%%%%%%%%%%%%%%%%%%%%%%%%%%%%%%%%%%
%
\begin{figure}
\center{\includegraphics[width=0.93\linewidth]{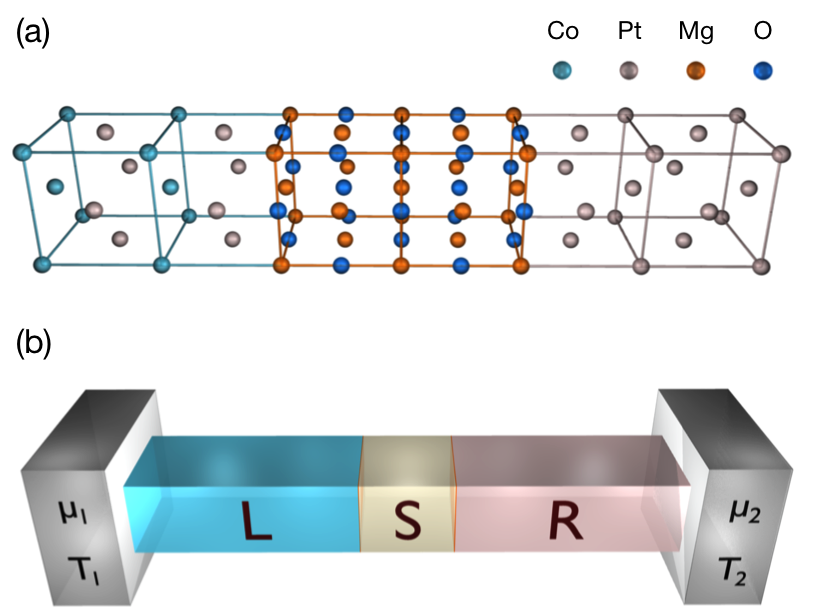}}
\caption{
\textbf{(a)}
Atomistic schematic of the \AS tunnel junction at the interfaces.  %The leads are grouped into principal layers, each consisting of two neighboring monolayers.  The principal layer scheme allows a Hamiltonian to retain a block-tridiagonal structure while admitting any range of interactions.
\textbf{(b)}
Schematic depicting the simplest geometry pertinent to the Landauer-B\"uttiker formalism.  Reflectionless contacts (depicted in gray) behave as thermal reservoirs, populated with carriers at a given Fermi Level $\mu$ and temperature $T$.  Carriers supplied from each contact proceed towards or away from the scattering region (S) via the left (L) and right (R) semi-infinite leads.
}
\label{fig:TheoreticalModel}
\end{figure}
%
%%%%%%%%%%%%%%%%%%%%%%%%%%%%%%%%%%%%%%%%%%%%%%%%%%%%%%
% End Figure Theoretical Model
%%%%%%%%%%%%%%%%%%%%%%%%%%%%%%%%%%%%%%%%%%%%%%%%%%%%%%

%%%%%%%%%%%%%%%%%%%%%%%%%%%%%%%%%%%%%%%%%%%%%%%%%%%%%%
% Figure Full 2DBZ Transport
%%%%%%%%%%%%%%%%%%%%%%%%%%%%%%%%%%%%%%%%%%%%%%%%%%%%%%
%
\begin{figure}
\center{\includegraphics[width=0.75\linewidth, trim = 0pt 0pt 0pt 0pt, clip]{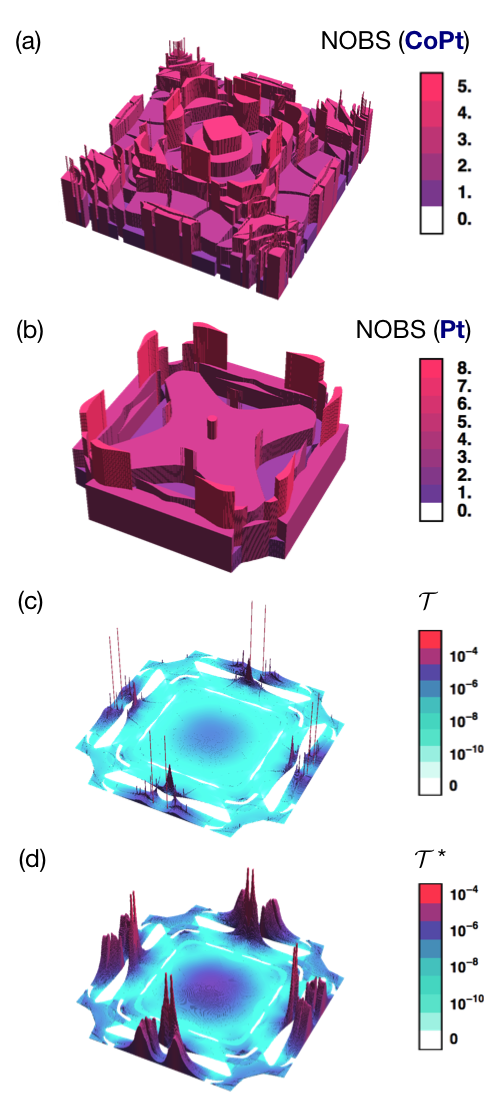}}
\caption{
Various transport quantities plotted over the 2DBZ at the Fermi energy.
\textbf{(a)}
The number of Bloch states (NOBS) traveling towards the scattering region in the CoPt lead ($\phi = 0\degree$).
\textbf{(b)}
The NOBS pertaining to the Pt lead.
\textbf{(c)}
The unitless transmission ($\T$) of the AMTJ, containing five MgO monolayers ($\phi = 0\degree$).  Sharp peaks known as ``hot spots" result from unrealistically clean interfaces.
\textbf{(d)}
The same transmission rescaled with hot-spots removed ($\T^*$).
}
\label{fig:Full2DBZTransport}
\end{figure}
%
%%%%%%%%%%%%%%%%%%%%%%%%%%%%%%%%%%%%%%%%%%%%%%%%%%%%%%
% End Figure Full 2DBZ Transport
%%%%%%%%%%%%%%%%%%%%%%%%%%%%%%%%%%%%%%%%%%%%%%%%%%%%%%

Throughout this work we use the following ratios
\begin{align}
	\text{Magnetoresistance (MR) Ratio} &= \Delta G / |G|_{max} 
	\label{eq:MRRatio} \\
	\text{Magneto-Seebeck (MS) Ratio} &= \Delta S / |S|_{max} 
	\label{eq:MSRatio}
\end{align}
to quantify the strengths of the aforementioned effects in both devices.  The numerators denote the greatest difference in conductance ($G$) or Seebeck coefficient ($S$) between any two magnetization directions.  The denominators represent the maximum absolute value of either quantity, yielding the so-called \emph{pessimistic} ratio.  We use these ratios to provide a consistent comparison between effects, and to avoid artificially high magneto-Seebeck ratios brought upon by vanishing Seebeck coefficients.

In regards to \OSNS, which exhibits the Tunneling Magnetoresistance (TMR) effect, we study the perpendicular-to-plane [001] parallel and antiparallel magnetization configurations.  On the other hand, \AS exhibits the Tunneling Anisotropic Magnetoresistance (TAMR) effect; for this system we rotate the free layer's magnetization from perpendicular-to-plane [001] to in-plane [100] over seven steps.  In analogy with the TMR and TAMR effects, the normal MTJ exhibits the Tunneling Magneto-Seebeck (TMS) effect while the anisotropic MTJ exhibits the Tunneling Anisotropic Magneto-Seebeck (TAMS) effect.

%\emph{Should we include the following paragraph?}
%In the following we explain our motivation for this study.  The magnetoresistance of tunnel junctions furnishes an electrically-measurable logic state for potential spintronics applications (such as MRAM).  Unfortunately, the exchange field produced by two ferromagnetic layers, in addition to the necessity for an antiferromagnetic pinning layer, limits the packing density and overall design simplicity of MTJ arrays.  While MTJs with single ferromagnetic contacts are desirable, their magnetoresistance ratios tend to be insufficient for commercial applications.  Rather than driving current through MTJs, one might produce temperature differences across them instead (for example, via heating currents).  In this case, the high and low states correspond to open circuit voltages induced across the tunnel junctions, the differences of which are ultimately given by the magneto-Seebeck ratio.  In the following we investigate which device structures yield the highest magnetoresistance and magneto-Seebeck ratios.

%%%%%%%%%%%%%%%%%%%%%%%%%%%%%%%%%%%%%%%%%%%%%%%%%%%%%%
% Figure Energy Dependent Transmission
%%%%%%%%%%%%%%%%%%%%%%%%%%%%%%%%%%%%%%%%%%%%%%%%%%%%%%
%
\begin{figure*}
\center{\includegraphics[width=1\linewidth, trim = 0pt 5pt 0pt 5pt, clip]{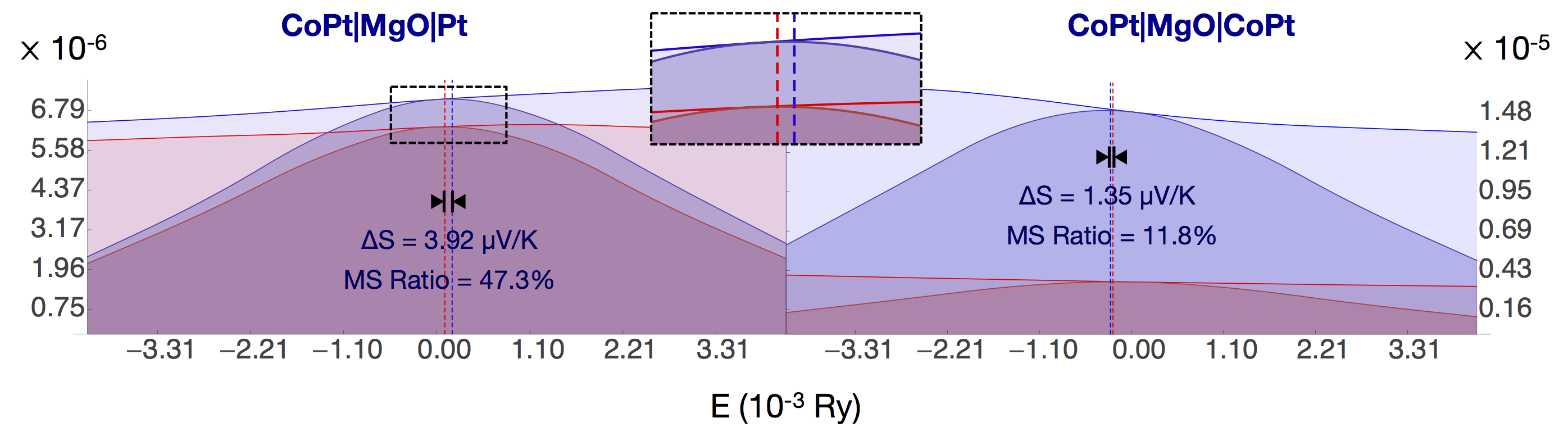}}
\caption[Visual representation of the magnetoresistance and magneto-Seebeck effects]{
Visual representation of the magnetoresistance and magneto-Seebeck effects at 300K for a four monolayer barrier thickness.  
Both $\TE$ (Eq.\ \ref{eq:LB}) and $\G$ (Eq.\ \ref{eq:GIntegrand}) are plotted versus energy for both devices.  
The lighter curves correspond to $\TE$ (unitless) while the darker curves represent $\G$ (Ry$^{-1}$).  
Only the magnetization directions yielding the high/low (blue/red) magnetoresistance states are shown.   
The energy-integral of $\G$ (the area under a darker curve) gives the conductance up to a factor of $e^2/h$ (Eq.\ \ref{eq:TotalConductance}).
The weighted center of $\G$ (a vertical dashed line) gives the Seebeck coefficient up to a factor of $e/hTG$ (Eq.\ \ref{eq:TotalSeebeckCoefficient}).
The zoomed panel pertains to \ASNS.
%
%Thus, regarding the darker curves, a large difference in their areas produces a strong magnetoresistance effect, while a large difference in their weighted centers yields a strong magneto-Seebeck effect.  
%
The actual difference in Seebeck coefficients pertaining to the blue and red dashed lines, as well as the corresponding magneto-Seebeck ratios, are shown for comparison.
%
%Note that $|S|_{max}$ normalizes the magneto-Seebeck ratio, defined in Eq.\ \ref{eq:MSRatio}. 
%
%As a result, \AS possesses the stronger magneto-Seebeck ratio, despite \OS having the greater difference in Seebeck coefficients.
}
\label{fig:EnergyDependentTransmission}
\end{figure*}
%
%%%%%%%%%%%%%%%%%%%%%%%%%%%%%%%%%%%%%%%%%%%%%%%%%%%%%%
% End Figure Energy Dependent Transmission
%%%%%%%%%%%%%%%%%%%%%%%%%%%%%%%%%%%%%%%%%%%%%%%%%%%%%%

We model each material system as a two-terminal device, consisting of a scattering center (MgO) and two semi-infinite leads (CoPt or Pt).  Our CoPt electrode consist of alternating monolayers of Co and Pt, ending with a Pt monolayer at the interface (Fig.\ \ref{fig:TheoreticalModel}a).  We subdivide the leads into principal layers, each consisting of two monolayers, so that our Hamiltonian retains a block-tridiagonal structure in the presence of next-nearest neighbor interactions.  The unit cell of each principal layer repeats periodically in-plane (perpendicular to transport), establishing a two-dimensional Brillouin Zone (2DBZ) per layer.  We assume that the corresponding wavevectors $\kp$ furnish good quantum numbers across the interfaces, enabling a common 2DBZ across each device.  To simulate a finite cross-sectional area, we enforce phase-repeating boundary conditions over some area in real space, constraining the available states in $\kp$-space to a minimum separation.  %Figure \ref{fig:DeviceModel} depicts the nuances of our device model discussed so far.

Our transport calculations utilize the Landauer-B\"uttiker formalism \cite{SivanImry1986, Buttiker}, in which transmission is computed via Green's Functions \cite{VelevButlerReview}.  We obtain all material Hamiltonians using the Slater-Koster tight-binding method \cite{SlaterKoster}, with parameters fitted to reproduce ab-initio electronic structure calculations \cite{Zemen201487, TanDFT}.  Within the ferromagnetic leads we use Stoner parameters to simulate the magnetization of the Co and Pt monolayers, the latter of which are slightly magnetized by proximity to Co.  We include spin-orbit coupling in all atoms to fully capture the magnetic transport-anisotropy.  Furthermore, we simulate interfacial strain between the leads and the sample via perturbations of hopping parameters.

The Landauer-B\"uttiker formalism expresses transport in terms of transmission probabilities obtained from a multi-dimensional scattering problem \cite{SivanImry1986, Buttiker}.  The scattering modes are eigenstates parameterized by the complex band structure of each lead.  For a given tunneling energy (E), transverse crystal momentum ($\kp$), and out-of-plane magnetization direction ($\phi$) in the free CoPt lead, the transmission function is given by
\begin{align}
	\T(\phi, E, \kp) &= Tr[\Gamma_L G^+ \Gamma_R G^-],
	\label{eq:Transmission}
\end{align}
\begin{align}
	\Gamma_{L(R)} &= i \big( \Sigma_{L(R)}^+ - \Sigma_{L(R)}^- \big),
	\label{eq:Gamma}
\end{align}
where $G$ represents the Green's function of the sample and $\Sigma$ gives the self-energy connecting a particular lead to the sample.  The subscripts $(L/R)$ denote the (left/right) leads, while the superscripts $(+/-)$ label the kind of Green's function (retarded/advanced) used to calculate that particular quantity.

Figure \ref{fig:Full2DBZTransport}c shows the transmission $\T(\kp)$ plotted over the 2DBZ for \AS with a five monolayer barrier thickness.  Sharp peaks known as ``hot spots" occur across the $\kp$-dependent transmission.  Theoretical studies of Fe-based MTJ's with crystalline MgO tunnel barriers show that these hot spots contribute negligibly to the calculated magnetoresistance ratios \cite{ButlerDeltaStates}.  In these cases, majority carriers belonging to the so-called $\Delta_1$ state dominate transport, overwhelming the contributions of hot spots found throughout the rest of the transmission.  %The $\Delta_1$ state (belonging to the complex band structure of MgO) possesses a U(1) rotational symmetry, which one can determine based on the symmetries of its constituent basis states.  Transmission, ultimately determined by wave-function matching at the interfaces, peaks when incoming states in the leads possess symmetries found in the scattering region.  In this way, MgO acts as a symmetry filter.

%The overall character of $\T(\kp)$ also depends on how one models their hopping parameters across the interfaces.  In our parameterization, we count 32 distinct hot spots in \AS with a five monolayer barrier thickness, when approximately one million $\kp$ points are included.  These 32 transmission values contribute over 6\% to the total Reimann sum; further analysis indicates that these regions require roughly one \emph{billion} points to fully resolve, rendering convergence impractical.  We assume such transmission resonances are unphysical since disorder introduces scattering across $\kp$ states, blurring $\T(\kp)$ and likely diminishing the contribution of hot spots.  

Enforcing phase-repeating boundary conditions constrains one to a maximum number of $\kp$ points (due to their minimum-allowed separation).  Our 2DBZ has an edge length of 1.2358 inverse Bohr radii; this implies that a cross-sectional area of 50 nm $\times$ 50 nm allows for $34,225$ $\kp$ points while 200 nm $\times$ 200 nm admits 552,049 $\kp$ points.  Between 66,049 and 263,169 $\kp$ points we find a compromise.  In this regime, $\T$ converges better than 2\% for both device structures and all magnetization directions, energies, and barrier widths -- if all hot spots are removed.  Following \cite{ZhangButlerRemoveHotSpots}, we remove hot spots in regions that would only converge well past a cutoff in the number of $\kp$ points (which we establish through finite-size considerations).

The sum of transmission probabilities over the 2DBZ gives the energy-dependent transmission $\TE$:
\begin{equation}
	\TE (\phi, E) =
	\sum_{\kp}
	\T(\phi, E, \kp).
	\label{eq:LB}
\end{equation}
To incorporate the effects of temperature, we populate available states with non-interacting electrons that obey Fermi-Dirac statistics, and neglect the contributions of inelastic phonons.  We assume that differences in temperature and electrochemical potential between the leads produce first-order variations in the Fermi-Dirac distribution with respect to energy, and no deviations in the transmission function (the linear-response limit).  In this approximation, we may express the conductance and Seebeck coefficient as
\begin{align}
	G (\phi, T) =
	\frac{e^2}{h}
	&\int
	\textbf{\textsf{G}} (\phi, E, T)
	dE 
	\label{eq:TotalConductance} \\
	S (\phi, T) =
	\frac{e}{h T G (\phi, T)}
	&\int
	( E - E_f )
	\textbf{\textsf{G}} (\phi, E, T)
	dE
	\label{eq:TotalSeebeckCoefficient}
\end{align}
respectively, where
\begin{equation}
	\textbf{\textsf{G}} (\phi, E, T) =
	\TE (\phi, E)
	\bigg( - \frac{\partial f }{\partial E} ( E, E_f, T ) \bigg)
	\label{eq:GIntegrand}
\end{equation}
is the energy-dependent transmission weighted by the derivative of the Fermi-Dirac distribution.  Note that all temperature-dependent quantities explicitly depend on the chosen Fermi level $E_f$.  

Eqs. $\ref{eq:TotalConductance}$ and $\ref{eq:TotalSeebeckCoefficient}$ provide a simple theoretical explanation for the origins of the magnetoresistance and magneto-Seebeck effects.  Whereas the conductance corresponds to the energy integral of $\textbf{\textsf{G}}$, the Seebeck coefficient corresponds to the geometric center of $\textbf{\textsf{G}}$ with respect to the Fermi level.  Thus, strong differences in the overall transmission, brought upon by rotating magnetization, lead to an appreciable magnetoresistance.  However, variations in the asymmetry of the energy-dependent transmission yield the magneto-Seebeck effect.

\begin{figure}
\center{\includegraphics[width=0.925\linewidth, trim = 0pt 0pt 0pt 0pt, clip]{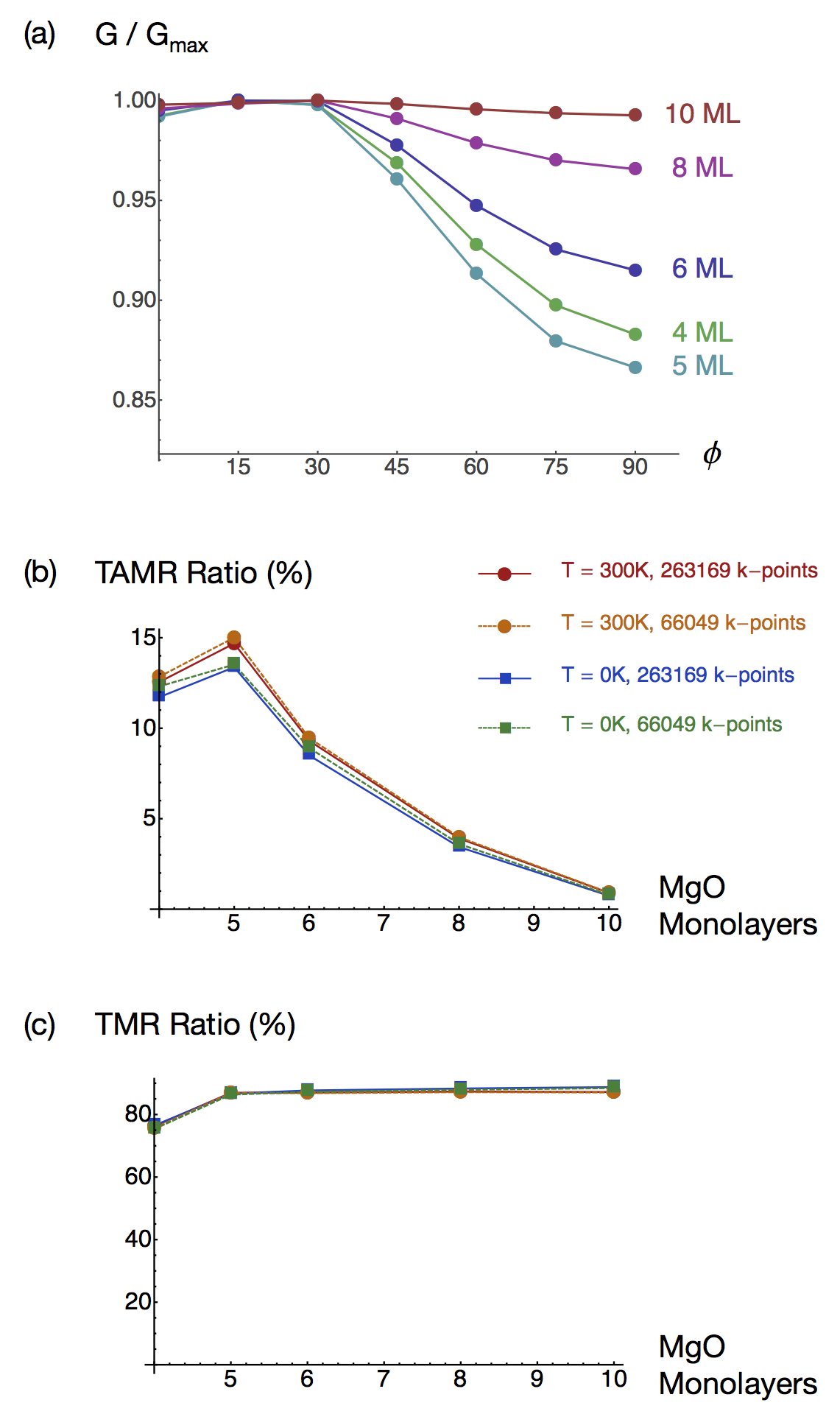}}
\caption{
Normalized TAMR curves shown for various numbers of MgO monolayers (ML) in the barrier (top panel).  TAMR (middle panel) and TMR (bottom panel) ratio versus barrier thickness, plotted for various temperatures and included numbers of $\kp$ points.  While the TAMR curve peaks at five MgO monolayers, the TMR curve saturates as barrier thickness increases.  The TMR ratios exceed the TAMR ratios by one order of magnitude.   
}
\label{fig:MRResults}
\end{figure}

\begin{figure}[h!]
\center{\includegraphics[width=1\linewidth, trim = 0pt 0pt 0pt 3pt, clip]{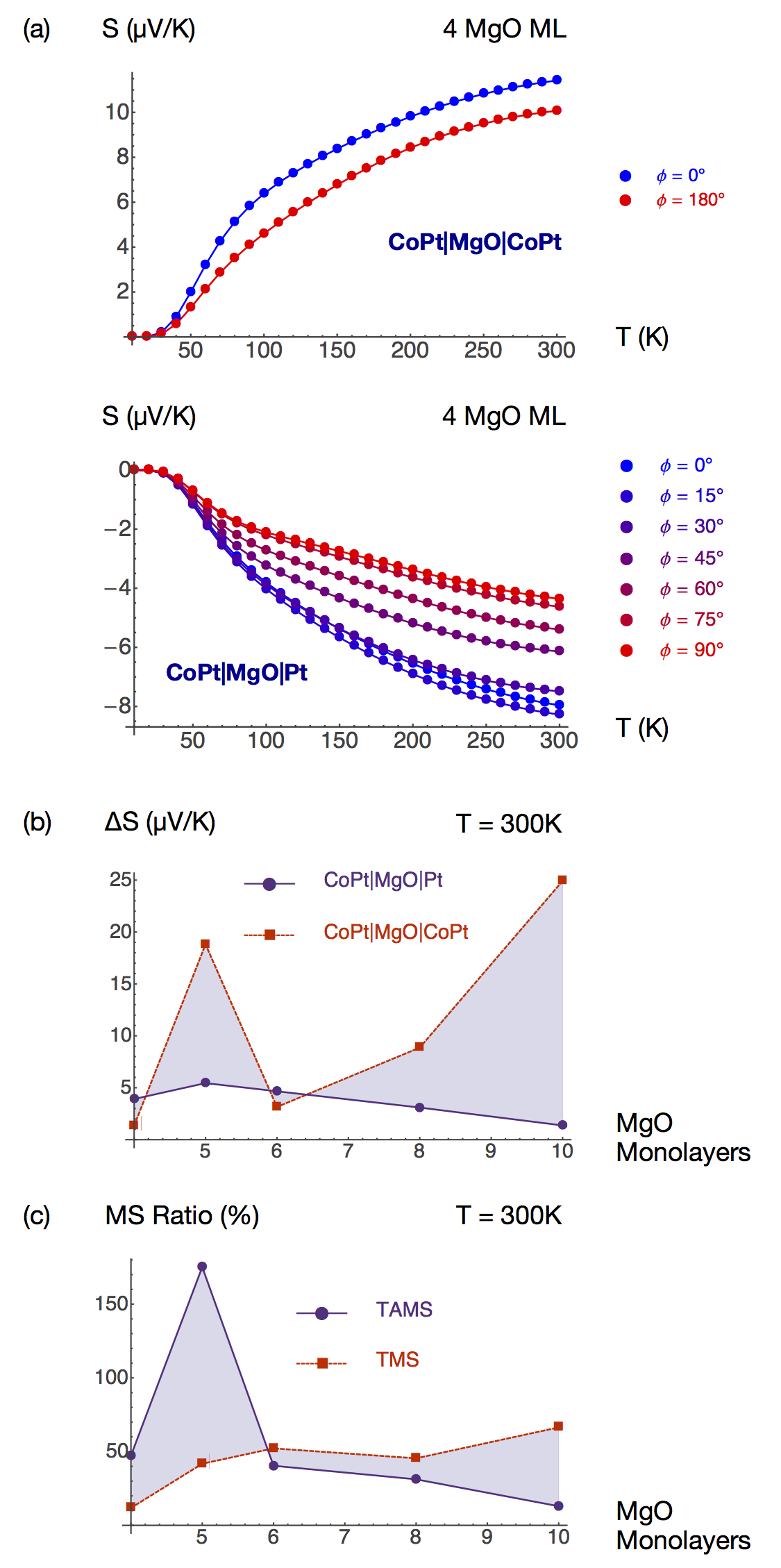}}
\caption{ 
\textbf{(a)} The Seebeck coefficient versus temperature, plotted for both devices and all magnetization directions (four MgO monolayers).  Both devices produce Seebeck coefficients of similar strength.  In addition, both devices produce comparable differences in Seebeck coefficient \textbf{(b)} and magneto-Seebeck ratios \textbf{(c)} for all barrier thicknesses simulated.  Furthermore, the TAMS ratio surpasses the TMS ratio for small barrier thicknesses, in contrast to the magnetoresistance ratios (Figs.\ \ref{fig:MRResults}b and \ref{fig:MRResults}c), which behave oppositely for all barrier thicknessess.
}
\label{fig:SeebeckandMSResults}
\end{figure}

We now discuss our results, beginning with the magnetoresistance effect.  Panel \ref{fig:MRResults}a displays normalized TAMR curves for various barrier thicknesses.  For barrier thicknesses greater than five MgO monolayers, the percentage difference between the high and low conductance states diminishes.  Panels \ref{fig:MRResults}b and \ref{fig:MRResults}c plot the simulated magnetoresistance ratios for various temperatures as a function of barrier thickness.  In accordance with panel \ref{fig:MRResults}a, the TAMR ratio peaks at five MgO monlayers (17.2\%) but decreases afterwards; however the TMR ratio saturates as barrier thickness increases (88.8\% at ten MgO monolayers).  In general the two devices produce magnetoresistance ratios roughly one order of magnitude apart.

%%%%%%%%%%%%%%%%%%%%%%%%%%%%%%%%%%%%%%%%%%%%%%%%%%%%%%
% Figure MR Results
%%%%%%%%%%%%%%%%%%%%%%%%%%%%%%%%%%%%%%%%%%%%%%%%%%%%%%
%
%
%%%%%%%%%%%%%%%%%%%%%%%%%%%%%%%%%%%%%%%%%%%%%%%%%%%%%%
% End Figure MR Results
%%%%%%%%%%%%%%%%%%%%%%%%%%%%%%%%%%%%%%%%%%%%%%%%%%%%%%

Moving now to the magneto-Seebeck effect, which encapsulates the main result of this letter, we direct the reader to Fig. \ref{fig:SeebeckandMSResults}.  Panel \ref{fig:SeebeckandMSResults}a shows the Seebeck coefficients plotted versus temperature for both devices (four MgO monolayers) and all magnetizations.  Here the normal and anisotropic MTJs yield Seebeck coefficients of similar strength.  Beyond four MgO monolayers, both devices produce comparable differences in Seebeck coefficient (panel \ref{fig:SeebeckandMSResults}b) and magneto-Seebeck ratios (panel \ref{fig:SeebeckandMSResults}c).  Furthermore, the magneto-Seebeck ratios of the anisotropic device \emph{surpass} those of the normal device for lower barrier thicknesses, peaking at absolute values of $68\%$ at $0$K and $175\%$ at $300$K.

%The normal MTJ seems better suited for recycling wasted heat (via the Seebeck effect).  However, one might consider the anisotropic MTJ for thermal logic applications (utilizing large magneto-Seebeck ratios), assuming that the benefits of a single ferromagnetic layer outweigh those of larger Seebeck coefficients.  

Unlike the conductance, the Seebeck coefficient may vanish, potentially causing the magneto-Seebeck ratio to diverge.  In general, small absolute values of $|S|_{max}$ produce artificially high magneto-Seebeck ratios; thus both the difference in Seebeck coefficient ($\Delta S$) and the magneto-Seebeck ratio factor into a device's performance.  In our case, although the Seebeck coefficients of some magnetization configurations vanish, $|S|_{max}$ (corresponding to $\phi$ = 90$\degree$ for the AMTJ and $\phi$ = 180$\degree$ for the MTJ) never does.

%%%%%%%%%%%%%%%%%%%%%%%%%%%%%%%%%%%%%%%%%%%%%%%%%%%%%%
% Figure TMS and TAMS vs Temperature and Barrier Thickness
%%%%%%%%%%%%%%%%%%%%%%%%%%%%%%%%%%%%%%%%%%%%%%%%%%%%%%
%
%
%%%%%%%%%%%%%%%%%%%%%%%%%%%%%%%%%%%%%%%%%%%%%%%%%%%%%%
% End Figure TMS and TAMS vs Temperature and Barrier Thickness
%%%%%%%%%%%%%%%%%%%%%%%%%%%%%%%%%%%%%%%%%%%%%%%%%%%%%%

In \ASNS, spin-polarized electrons leaving the ferromagnetic layer tunnel into a region with no spin preference; thus the single ferromagnetic layer alone controls the magnetic transport anisotropy (through spin-orbit coupling).  However, in \OSNS, spins polarized by the first layer tunnel into a receiving layer with either strong (parallel configuration) or weak (antiparallel configuration) preference for their spin, providing a stronger spin-filter irrespective of spin-orbit coupling.  Thus, anisotropic MTJs often yield lower magnetoresistance ratios than normal MTJs.  Thermally-induced voltages appear to behave differently.  The magneto-Seebeck effect stems from changes in the asymmetry of $\TE(E)$ brought upon by rotating magnetization.  In principle, such variations in asymmetry need not be connected to variations in the overall transmission.  In agreement with this assumption, our results indicate that \AS possesses thermal transport anisotropy similar or better than \OSNS.

%%%%%%%%%%%%%%%%%%%%%%%%%%%%%%%%%%%%%%%%%%%%%%%%%%%%%%
% Conclusion
%%%%%%%%%%%%%%%%%%%%%%%%%%%%%%%%%%%%%%%%%%%%%%%%%%%%%%

To conclude, we have demonstrated that magnetic tunnel junctions possessing a single ferromagnetic layer can produce magneto-Seebeck ratios exceeding those of normal MTJs.  This behavior provides a sharp contrast to that of the magnetoresistance.  We performed coherent transport calculations simulating the magnetoresistance and magneto-Seebeck effects in \OS and \AS magnetic tunnel junctions.  The anisotropic MTJ yields magneto-Seebeck ratios comparable or better to those of the normal MTJ, reaching absolute values of $175\%$ at room temperature.   Thus we find that exploiting spin-orbit coupling in MTJs with a single ferromagnetic contact can lead to enhanced magnetic-transport anisotropies.

\begin{acknowledgements}
The authors gratefully acknowledge Jan Ma\v sek for helpful discussions. This work was supported by the National Science Foundation through the Materials Research Science and Engineering Center at The Ohio State University (DMR-0820414), EPSRC grant no.\ EP/H029257/1, the EU European Research Council (ERC) advanced grant no.\ 268066, the Ministry of Education of the Czech Republic grant no.\ LM2011026, the Grant Agency of the Czech Republic grant no.\ 14-37427G, the Academy of Sciences of the Czech Republic Praemium Academiae, and the Alexander Von Humboldt Foundation.
\end{acknowledgements}

%%%%%%%%%%%%%%%%%%%%%%%%%%%%%%%%%%%%%%%%%%%%%%%%%%%%%%
% End Main Body
%%%%%%%%%%%%%%%%%%%%%%%%%%%%%%%%%%%%%%%%%%%%%%%%%%%%%%

%%%%%%%%%%%%%%%%%%%%%%%%%%%%%%%%%%%%%%%%%%%%%%%%%%%%%%
% Begin Bibliography
%%%%%%%%%%%%%%%%%%%%%%%%%%%%%%%%%%%%%%%%%%%%%%%%%%%%%%

\bibliographystyle{plain}
\bibliography{References}

\begin{thebibliography}{10}

\bibitem{BauerSpinCaloritronics}
G.~E.~W. Bauer, E.~Saitoh, and B.~J. van Wees.
\newblock {\em Nature Materials}, 11:391--399, May 2012.

\bibitem{TMSExperiment2}
A.~Boehnke, M.~Walter, N.~Roschewsky, T.~Eggebrecht, V.~Drewello, K.~Rott,
  M.~Munzenberg, A.~Thomas, and G.~Reiss.
\newblock {\em Review of Scientific Instruments}, 84(6):--, 2013.

\bibitem{TMSExperiment1}
T.~Bohnert, V.~Vega, A-K Michel, V.~M. Prida, and K.~Nielsch.
\newblock {\em Applied Physics Letters}, 103(9):--, 2013.

\bibitem{ButlerDeltaStates}
W.~H. Butler, X.-G. Zhang, T.~C. Schulthess, and J.~M. MacLaren.
\newblock {\em Phys. Rev. B}, 63:054416, Jan 2001.

\bibitem{Buttiker}
M.~Buttiker.
\newblock {\em Phys. Rev. B}, 46:12485--12507, Nov 1992.

\bibitem{MagnetoSeebeckTheoryAbInitio2}
M~Czerner and C~Heiliger.
\newblock {\em Journal of Applied Physics}, 111(7), 2012.

\bibitem{MagnetoSeebeckTheoryAbInitio1}
C~Heiliger, C~Franz, and M~Czerner.
\newblock {\em Phys. Rev. B}, 87:224412, Jun 2013.

\bibitem{LiebingMagnetothermopower}
N.~Liebing, S.~Serrano-Guisan, K.~Rott, G.~Reiss, J.~Langer, B.~Ocker, and
  H.~W. Schumacher.
\newblock {\em Phys. Rev. Lett.}, 107:177201, Oct 2011.

\bibitem{MagnetoSeebeckTheoryBoltzmann2}
C~Lopez-Mon's, A~Matos-Abiague, and J~Fabian.
\newblock {\em arXiv:1309.3463}, 2013.

\bibitem{TAMSExperiment2}
M.~Magnuson, M.~Mattesini, N.~V. Nong, P.~Eklund, and L.~Hultman.
\newblock {\em Phys. Rev. B}, 85:195134, May 2012.

\bibitem{NaydenovaThermopower}
Ts. Naydenova, P.~D\"urrenfeld, K.~Tavakoli, N.~P\'egard, L.~Ebel, K.~Pappert,
  K.~Brunner, C.~Gould, and L.~W. Molenkamp.
\newblock {\em Phys. Rev. Lett.}, 107:197201, Oct 2011.

\bibitem{SivanImry1986}
U.~Sivan and Y.~Imry.
\newblock {\em Phys. Rev. B}, 33:551--558, Jan 1986.

\bibitem{SlaterKoster}
J.~C. Slater and G.~F. Koster.
\newblock {\em Phys. Rev.}, 94:1498--1524, Jun 1954.

\bibitem{TanDFT}
Y.~Tan, M.~Povolotskyi, T.~Kubis, Y.~He, Z.~Jiang, G.~Klimeck, and T.~B.
  Boykin.
\newblock {\em Journal of Computational Electronics}, 12(1):56--60, 2013.

\bibitem{TAMSExperiment1}
G.~D. Tang, H.~H. Guo, T.~Yang, D.~W. Zhang, X.~N. Xu, L.~Y. Wang, Z.~H. Wang,
  H.~H. Wen, Z.~D. Zhang, and Y.~W. Du.
\newblock {\em Applied Physics Letters}, 98(20):--, 2011.

\bibitem{VelevButlerReview}
Julian Velev and William Butler.
\newblock {\em Journal of Physics: Condensed Matter}, 16(21):R637, 2004.

\bibitem{WalterMagnetoSeebeck}
M.~Walter, J.~Walowski, V.~Zbarsky, M.~Munzenberg, M.~Schafers, D.~Ebke,
  G.~Reiss, A.~Thomas, P.~Peretzki, M.~Seibt, J.~S. Moodera, M.~Czerner,
  M.~Bachmann, and C.~Heiliger.
\newblock {\em Nature Materials}, 10:742--746, Oct 2011.

\bibitem{Zemen201487}
J.~Zemen, J.~Ma\v{s}ek, J.~Ku\v{c}era, J.A. Mol, P.~Motloch, and T.~Jungwirth.
\newblock {\em Journal of Magnetism and Magnetic Materials}, 356(0):87 -- 94,
  2014.

\bibitem{ZhangButlerRemoveHotSpots}
X.-G. Zhang and W.~H. Butler.
\newblock {\em Phys. Rev. B}, 70:172407, Nov 2004.

\end{thebibliography}

%%%%%%%%%%%%%%%%%%%%%%%%%%%%%%%%%%%%%%%%%%%%%%%%%%%%%%
% End Bibliography
%%%%%%%%%%%%%%%%%%%%%%%%%%%%%%%%%%%%%%%%%%%%%%%%%%%%%%

\end{document}